\documentclass[12pt]{article}
\usepackage{graphicx}
\usepackage{subfigure}
\usepackage{amsmath,amsthm,amscd,amssymb}

\begin{document}

\title{Social networks that matter: \texttt{Twitter} under the microscope}
\author{Bernardo A. Huberman$^{1}$, Daniel M. Romero$^{1,2}$ and Fang Wu$^{1}$\\ \small $^{1}$Social Computing
Lab, HP Laboratories, Palo Alto, CA 94304 \\ \small $^{2}$Cornell University, Ithaca, NY 14850}

\maketitle

\begin{abstract}
Scholars, advertisers and political activists see massive online social
networks as a representation of social interactions that can be used to study
the propagation of ideas, social bond dynamics and viral marketing, among others.
But the linked structures of social networks do not reveal actual interactions among
people. Scarcity of attention and the daily rythms of life and work makes people
default to interacting with those few that matter and that reciprocate their
attention.  A study of social interactions within \texttt{Twitter} reveals that the driver of usage is a sparse and hidden network of connections underlying the ``declared'' set of friends and followers. 
\end{abstract}
\pagebreak

Social networks, a very old and pervasive mechanism for mediating distal
interactions among people, have become prevalent in the age of the Web.
With interfaces that allow people to follow the lives of friends, acquaintances and
families, the number of people on social networks has grown exponentially since the
turn of this century. \texttt{Facebook}, \texttt{LinkedIn} and \texttt{MySpace}, to give a few examples, contain
millions of members who use these networks for keeping track of each other, find
experts and engage in commercial transactions when needed \cite{convergence}. Furthermore, commercial
enterprises try to exploit them for marketing purposes, as they provide a ready made
medium for propagating recommendations through people with similar
interests \cite{viral}.

On the academic side, a large body of knowledge has accumulated on the formation and
dynamics of these networks, fueled by the easy availability of data and the
regularities found in the statistical distribution of nodes and links within these
networks \cite{friends, weakties, teen, students, methods, wired}.

While the standard definition of a social network embodies the notion of all the
people with whom one shares a social relationship, in reality people interact with
very few of those ``listed'' as part of their network. One important reason behind
this fact is that attention is the scarce resource in the age of the web. Users
faced with many daily tasks and large number of social links default to interacting
with those few that matter and that reciprocate their attetention. For example, a
recent study of \texttt{Facebook} showed that users only poke and message a small number of
people while they have a large number of declared friends \cite{facebook}. And a
casual search through recent calls made through any mobile phone usually reveals
that a small percentage of the contacts stored in the phone are frequently contacted by the user.

These initial observations suggest a systematic investigation into the nature of the
social networks that actually matter to people. By networks that matter we mean
those networks that are made out of the pattern of interactions that people have
with their friends or acquaintances, rather than constructed from a list of all the
contacts they may decide to declare.

In order to find out how relevant a list of ``friends'' is to members of the network,
we collected and analyzed a large data set from the \texttt{Twitter} social network.
\texttt{Twitter.com} is a online social network used by millions of people around the
world  to stay connected to their friends, family members and coworkers through
their computers and mobile phones. The interface allows users to post short messages (up to 140
characters) that can be read by any other \texttt{Twitter} user. Users declare the people
they are interested in following, in which case they get notified when that person
has posted a new message. A user who is being followed by another user does not
necessarily have to reciprocate by following them back, which makes the links of the
\texttt{Twitter} social network directed.

For each user of \texttt{Twitter} in our data set we obtained the number of \emph{followers} and \emph{followees} (people followed by a user) the user
has declared, along with the content and datestamp of all his posts.\footnote{\texttt{Twitter} only displays up to 3201 updates per user so we only have the complete set of updates for users who have posted  3200 or less updates.  A very
small set of users showed 3201 updates so we have the complete set for about 99.6\%
of all the users.} Our data set consisted of a total of 309,740 users, who on average posted 255 posts, had 85 followers, and followed 80 other users. Among the 309,740 users only 211,024 posted at least twice. We call them the \emph{active users}. We also define the
\emph{active time} of an active user by the time that has elapsed between his first and last
post. On average, active users were active for 206 days.

\texttt{Twitter} users are able to post direct and indirect updates. Direct posts are used
when a user aims her update to a specific person, whereas indirect updates are used
when the update is meant for anyone that cares to read it. Even though direct
updates are used to communicate directly with a specific person, they are public and
anyone can see them. Often times two or more users will have conversations by
posting updates directed to each other. Around 25.4\% of all posts are directed,
which shows that this feature is widely used among \texttt{Twitter} users.

We are interested in finding out how many people each user communicates directly
with through \texttt{Twitter}. We define a user's \emph{friend} as a person whom the user has
directed at least two posts to. Using this definition we were able to find out how
many friends each user has and compare this number with the number of followers and
followees they declared.

%%%%%%%%%%%%%%%%%%%%%
%%FollowersVsTotalPosts
%%%%%%%%%%%%%%%%%%%%%
{\begin{figure}
\begin{center}
\includegraphics[width = 1 \textwidth, height =.5 \textwidth]
{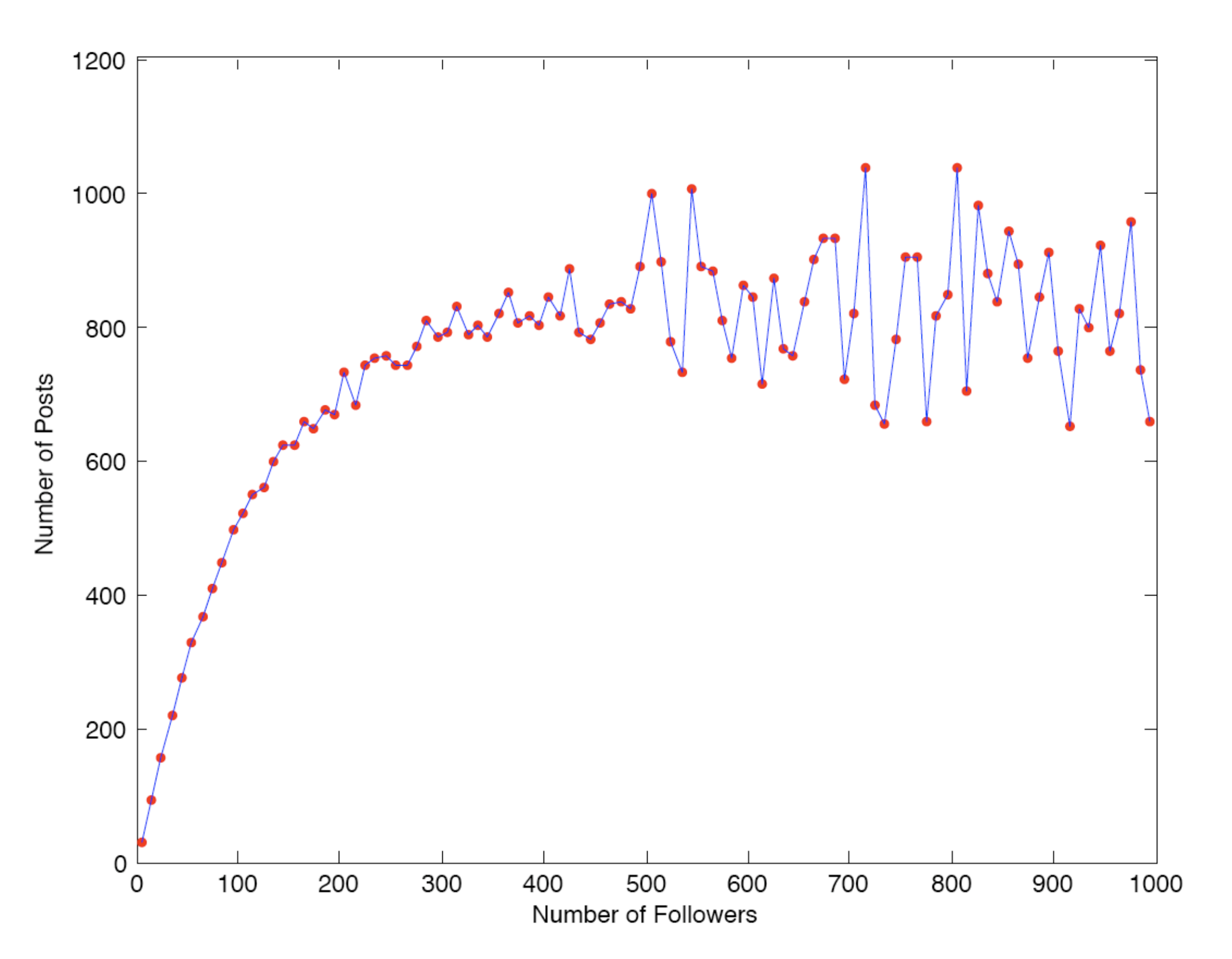} \end{center} \caption{\small{\textbf{Number of posts as
a function of the number of followers.
The number of posts initially increases as the number of followers increases but it
eventually saturates.}}} \label{FollowersVsTotalPosts} \end{figure}

%%%%%%%%%%%%%%%%%%%%%
%%FriendsVsTotalPosts
%%%%%%%%%%%%%%%%%%%%%
{\begin{figure}
\begin{center}
\includegraphics[width = 1 \textwidth, height =.5 \textwidth]
{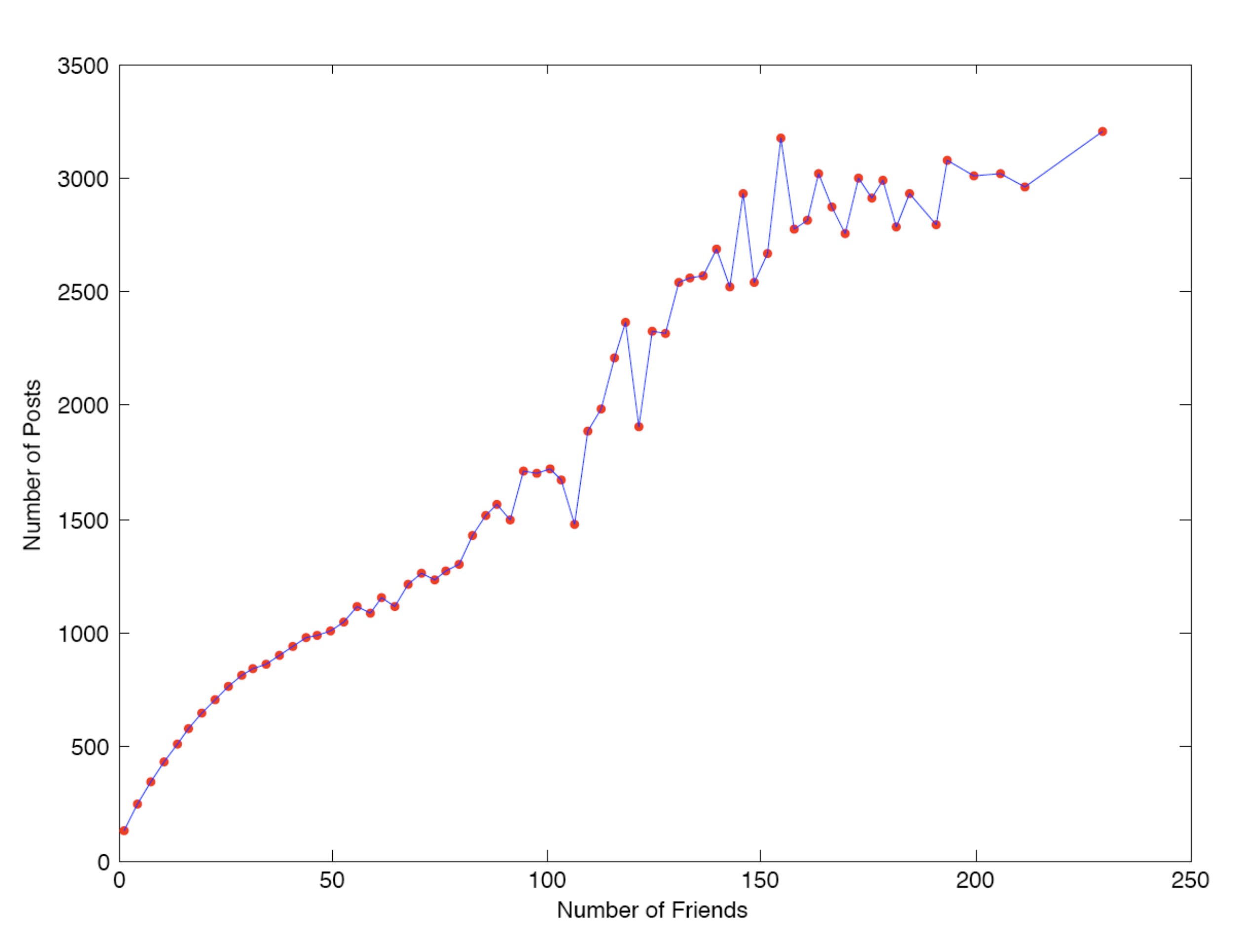} \end{center} \caption{\small{\textbf{Number of posts as a
function of the number of friends. The number of posts increases as the number of
friends increases, reaching 3200 without saturating.}}} \label{FriendsVsTotalPosts}
\end{figure}

Based on our previous finding about the role of attention in eliciting productivity
within a social network \cite{crowdsourcing}, we conjecture that the users who
receive attention from many people will post more often than users who receive
little attention.
Therefore we expect that users with more followers and friends will be more active
at posting than those with a small number of followers and friends.  Figures
\ref{FollowersVsTotalPosts} and \ref{FriendsVsTotalPosts} show that indeed the total
number of posts increases with both the number of followers and friends.
However, as figure \ref{FollowersVsTotalPosts} shows, the number of total posts
eventually saturates as a function of the number of followers. This implies that
users with a large number of followers are not necessarily those with very large
number of total posts. On the other hand, the number of total posts does not
saturate as a function of number of friends, as seen on figure
\ref{FriendsVsTotalPosts}. Rather, the number of updates increases until it reaches
a maximum point of 3201. This suggests that in order to predict how active a \texttt{Twitter}
user is, the number of friends is a more accurate signal than the number of his
followers.

This implies that to assess the size of the social network that matters we need to consider those people who actually communicate though direct messages with each other,
as opposed to the network created by the declared followers and followees.

Having shown that the number of friends is the actual driver of \texttt{Twitter} user's
activity, we compared it with the number of followees the users declare.
We define $\delta$ as the number of friends a user has, divided by the number of
followees she declared. Since 98.8\% of the users have fewer friends than followees,
almost all the $\delta$ values are less than 1. Figure \ref{FriendsHistogram} shows
a histogram of the $\delta$ values. As we can see most users have a $\delta$ value
less than .1, with the number of users with a $\delta$ close to 1 extremely small.
The average of the $\delta$ values is 0.13 and the median is 0.04. This indicates
that the number of friends users have is very small compared to the number of people
they actually follow. Thus, even though users declare that they follow many people
using \texttt{Twitter}, they only keep in touch with a small number of them. Hence, while the
social network created by the declared followers and followees appears to be very
dense, in reality the more influential network of friends suggests that the social
network is sparse.

%%%%%%%%%%%%%%%%%%%%%
%%FriendsHistogram
%%%%%%%%%%%%%%%%%%%%%
{\begin{figure}
\begin{center}
\includegraphics[width = 1 \textwidth, height =.5 \textwidth] {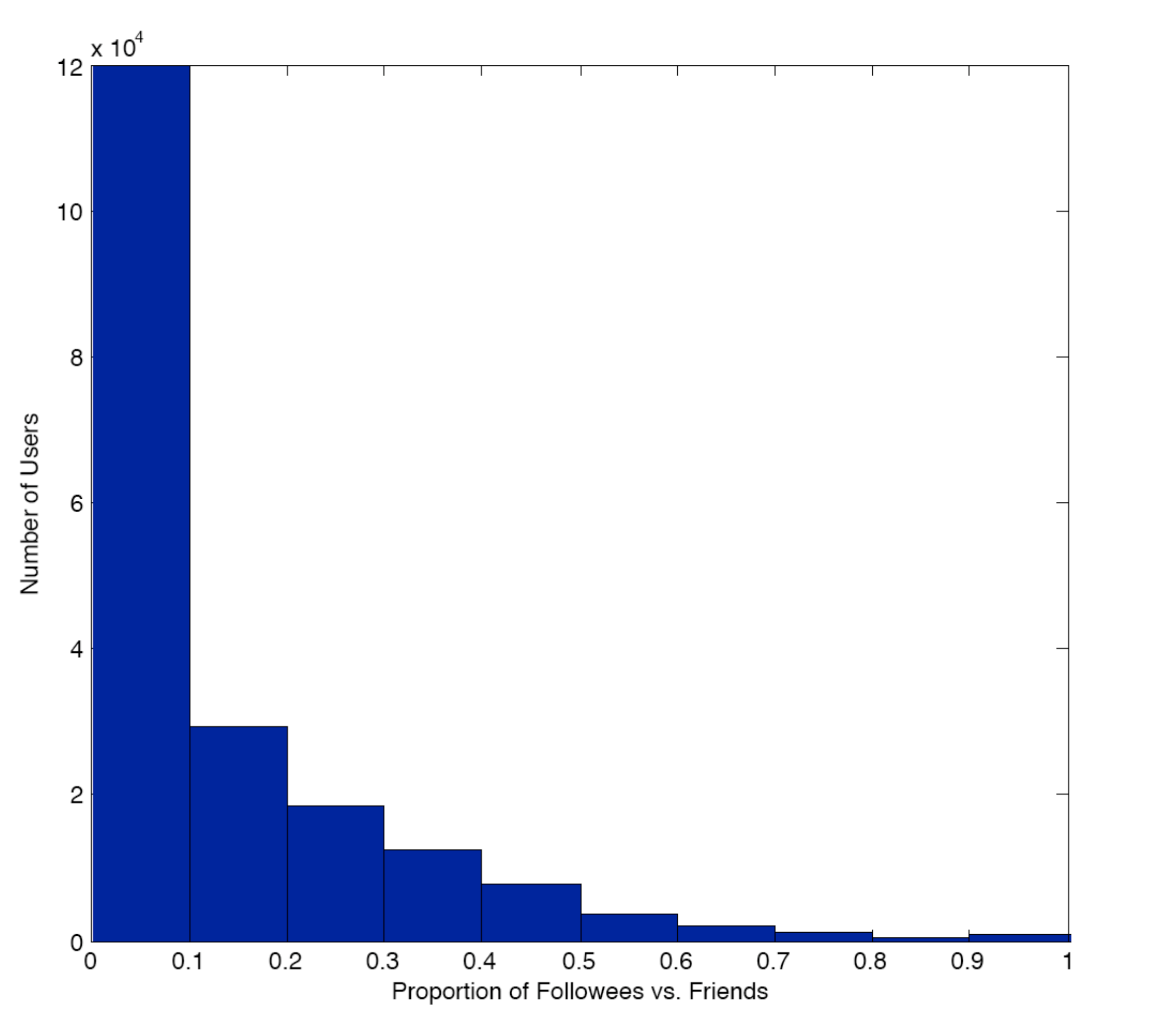}
\end{center} \caption{\small{\textbf{Histogram of contributor's number of friends
divided by the number of followees. Most users have a very small number of friends
compared to the number of followees they declared.}}} \label{FriendsHistogram}
\end{figure}

Another interesting aspect is to consider how the number of friends and the $\delta$ values 
change as the number of followees increases. Figures \ref{FolloweesVsFriends} and
\ref{ProportionVsFollowees} show that even though the number of friends initially
increases as the number of followees increases, after a while the number of friends
starts to saturate and stays nearly constant. This trend can be explained by the
fact that the cost of declaring a new followee is very low compared to the cost of
maintaining a friends (i.e.~exchanging directed messages with other users). Hence,
the number of people a user actually communicates with eventually stops increasing
while the number of followees can continue to grow indefinitely.

%%%%%%%%%%%%%%%%%%%%%
%%FolloweesVsFriends
%%%%%%%%%%%%%%%%%%%%%
{\begin{figure}
\begin{center}
\includegraphics[width = 1 \textwidth, height =.5 \textwidth]
{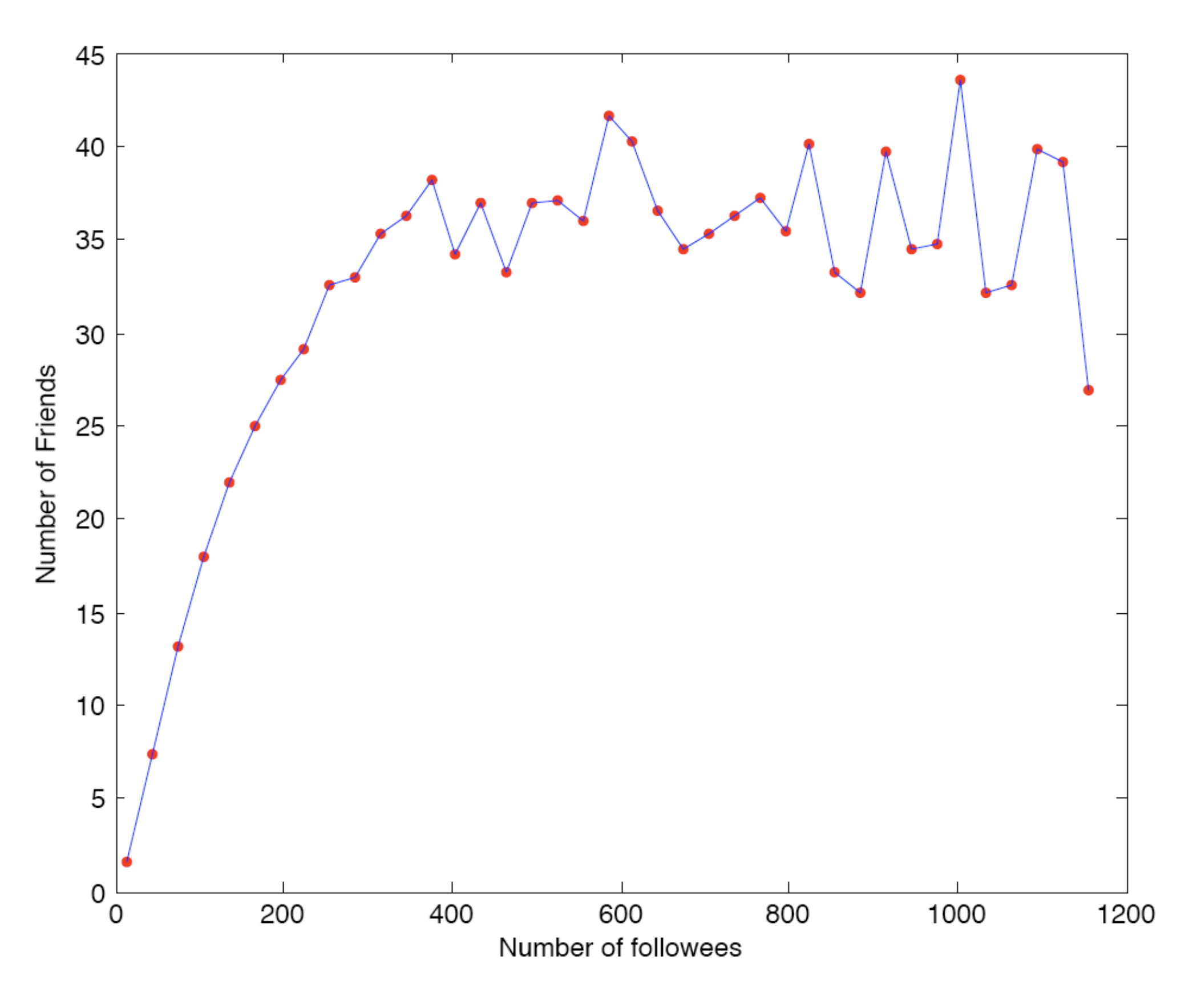} \end{center} \caption{\small{\textbf{Number of friends as a
function of the number of followees.
The total number of friends saturates while the number of followees keeps
growing due to the minimal effort required to add a followee.}}}
\label{FolloweesVsFriends} \end{figure}

%%%%%%%%%%%%%%%%%%%%%
%%ProportionVsFollowees
%%%%%%%%%%%%%%%%%%%%%
{\begin{figure}
\begin{center}
\includegraphics[width = 1 \textwidth, height =.5 \textwidth]
{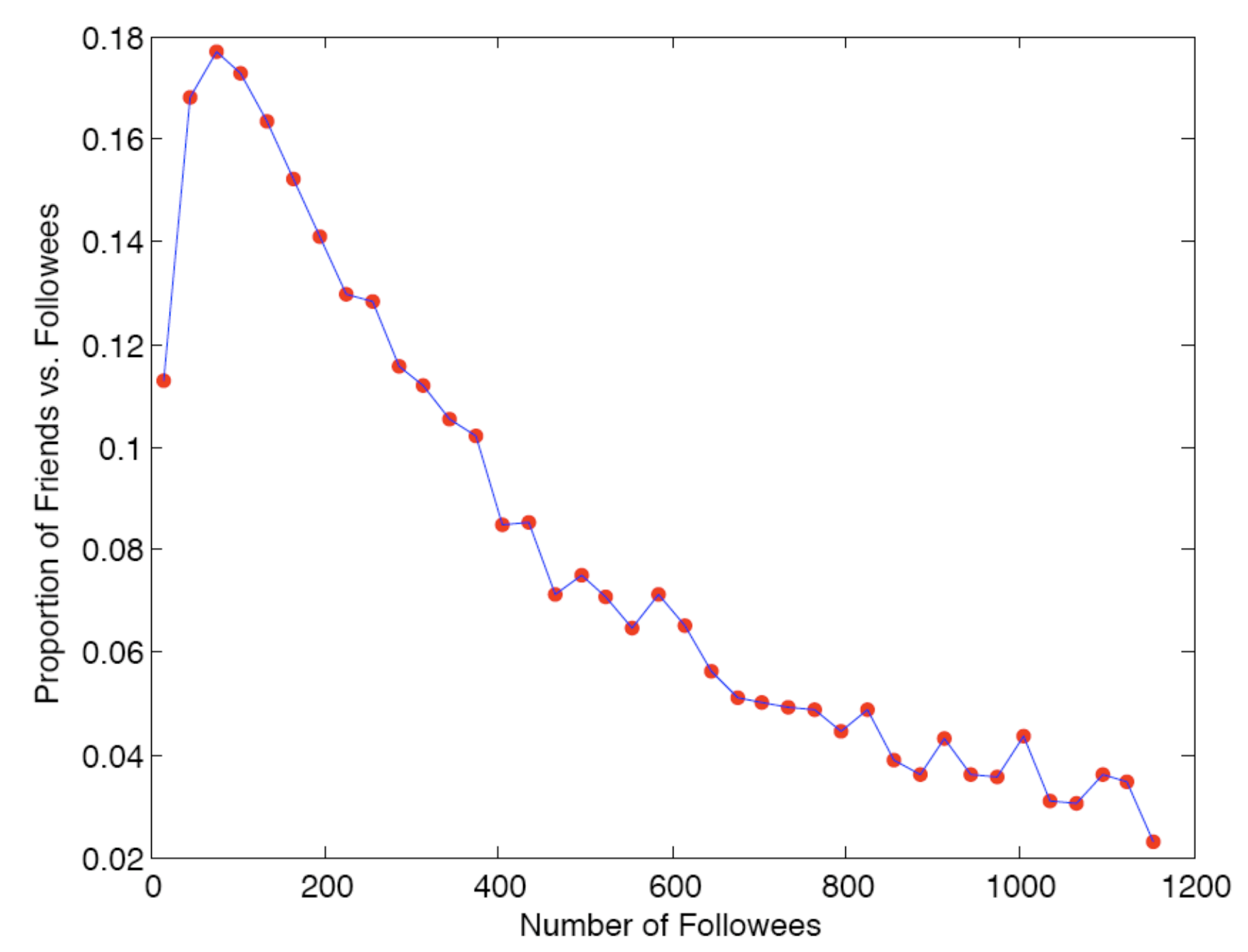} \end{center} \caption{\small{\textbf{Proportion of
friends vs.~followees as a function of followers. It initially increases but
rapidly approaches zero as the number of followees increases.}}}
\label{ProportionVsFollowees} \end{figure}

In conclusion, even when using a very weak definition of ``friend'' (i.e.~anyone who a
user has directed a post to at least twice) we find that \texttt{Twitter} users have a very
small number of friends compared to the number of followers and followees they
declare. This implies the existence of two different networks: a very dense one made
up of followers and followees, and a sparser and simpler network of actual friends.
The latter proves to be a more influential network in driving \texttt{Twitter} usage since
users with many actual friends tend to post more updates than users with few actual
friends. On the other hand, users with many followers or followees post updates more
infrequently than those with few followers or followees.

Many people, including scholars, advertisers and political activists, see online
social networks as an opportunity to study the propagation of ideas, the formation
of social bonds and viral marketing, among others. This view should be tempered by
our findings that a link between any two people does not necessarily imply an
interaction between them.  As we showed in the case of \texttt{Twitter}, most of the links
declared within \texttt{Twitter} were meaningless from an interaction point of view. Thus the
need to find the hidden social network; the one that matters  when trying to rely on
word of mouth to spread an idea, a belief, or a trend.

%%%%%%%%%%%%%%%%%%%%%
%%Graph of Networks
%%%%%%%%%%%%%%%%%%%%
\begin{figure}
\centering
\subfigure[\tiny{\textbf{All links are declared followees and the red links are
actual friends.}}]{ \includegraphics[scale=.35]{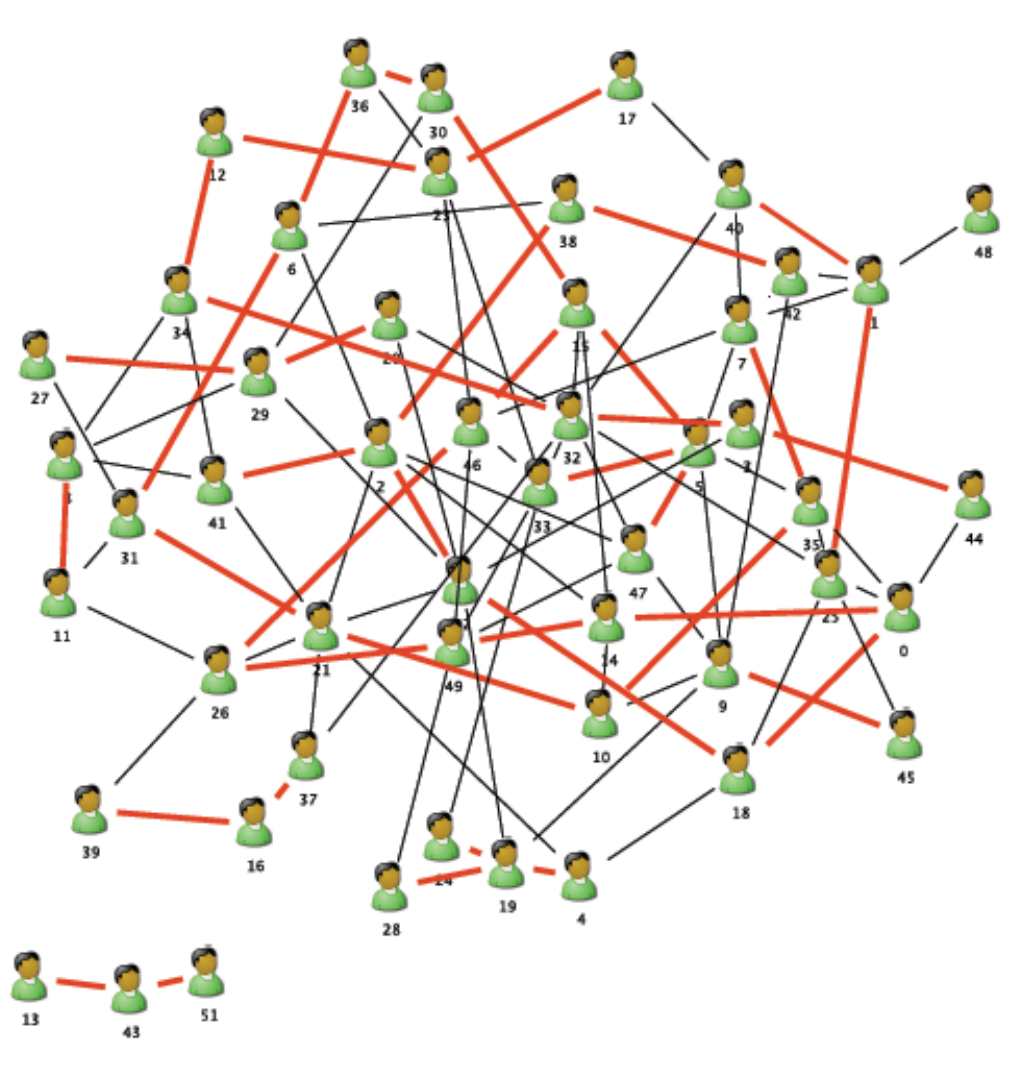}
\label{fig:subfig1}
}
\subfigure[\tiny{\textbf{After removing the black links and reorganizing the
network look simpler than before. This is the hidden network that matters the
most.}}]{ \includegraphics[scale=.35]{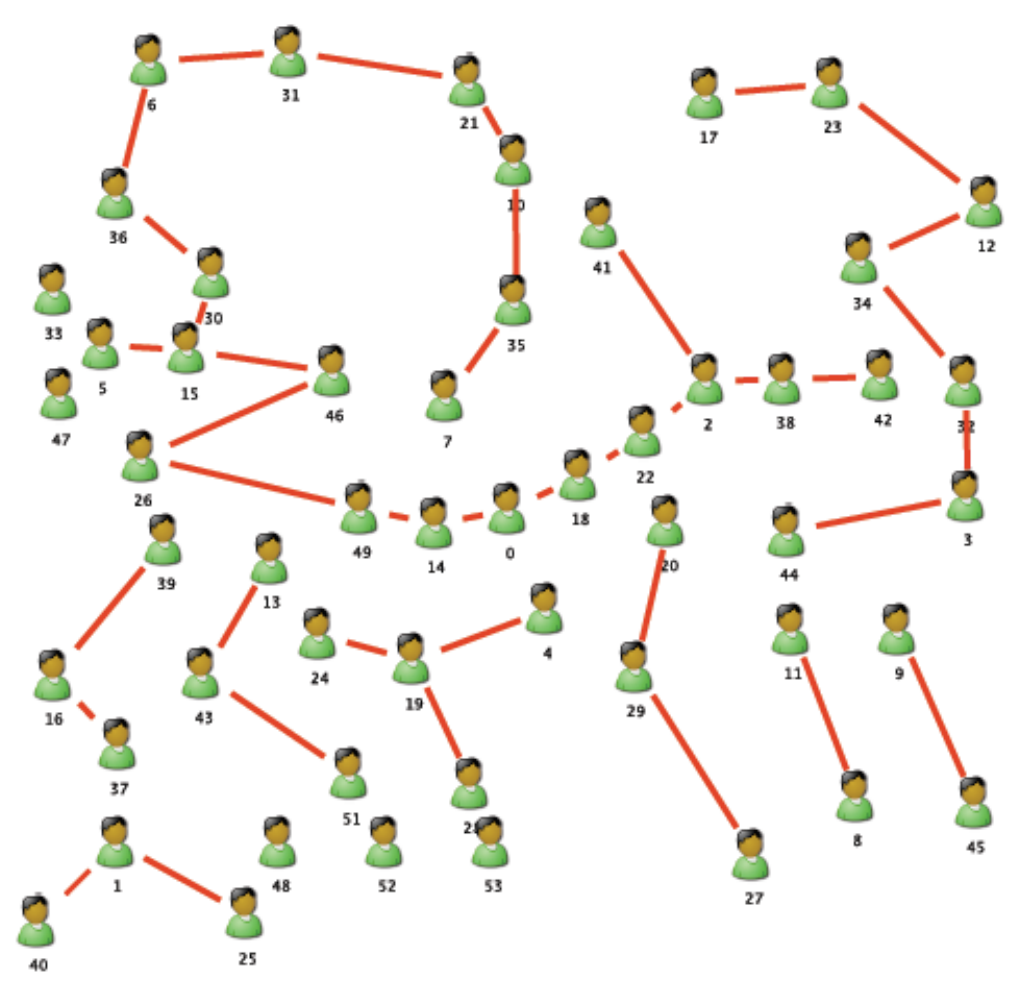}
\label{fig:subfig2}
}
\end{figure}

\end{document}